\documentclass[pre,twocolumn,twoside,showpacs,byrevtex,superscriptaddress]{revtex4}

\usepackage{array}

\newcommand{\PreserveBackslash}[1]{\let\temp=\\#1\let\\=\temp}

\usepackage{graphicx,epsfig,verbatim,enumerate}
\usepackage{amssymb,amsmath}
\usepackage{ifthen}
\newboolean{twocolswitch}

\usepackage{color}

\newcommand{\tavg}[1]{\langle#1\rangle}

\newcommand{\sindex}[1]{}
\newcommand{\nindex}[1]{}

\newcommand{\etal}{\textit{et al.}}
\newcommand{\www}[1]{\url{#1}}

\newcommand{\req}[1]{(\ref{#1})}
\newcommand{\Req}[1]{Eq.~(\ref{#1})}

\newcommand{\Ptrig}{P_{\rm trig}}

\newcommand{\veck}{\vec{k}}

\newcommand{\bidmark}{\rm u}
\newcommand{\inmark}{\rm i}
\newcommand{\outmark}{\rm o}

\newcommand{\undmark}{\rm und}
\newcommand{\dirmark}{\rm dir}

\newcommand{\kbid}{k_{\bidmark}}
\newcommand{\kin}{k_{\inmark}}
\newcommand{\kout}{k_{\outmark}}

\newcommand{\jbid}{j_{\bidmark}}
\newcommand{\jin}{j_{\inmark}}

\newcommand{\Probbid}{P^{(\rm \bidmark)}}
\newcommand{\Probdir}{P^{(\rm \dirmark)}}
\newcommand{\Probin}{P^{(\rm \inmark)}}
\newcommand{\Probout}{P^{(\rm \outmark)}}

\newcommand{\edgetype}{\veck\, |\, \veck'}

\newcommand{\probfc}{Q}

\newcommand{\indrspfn}{\mathcal{B}}
\newcommand{\rspfn}{B}

\newcommand{\infparam}{\beta}

\newcommand{\erdosrenyi}{Erd\H{o}s-R\'{e}nyi}

\newcommand{\toyund}{\tau_{\undmark}}
\newcommand{\toydir}{\tau_{\dirmark}}

\newcommand{\tinfbid}{\theta_{\veck_{1},\infty}^{(\bidmark)}}
\newcommand{\tinfin}{\theta_{\veck_{1},\infty}^{(\inmark)}}

\newcommand{\Quk}{\probfc^{(\rm u)}_{\veck_1}}
\newcommand{\Qok}{\probfc^{(\rm o)}_{\veck_1}}

\setboolean{twocolswitch}{true}

\begin{document}

\title{
  Exact solutions for social and biological contagion models on 
mixed directed and undirected, degree-correlated random networks 

}

\author{
\firstname{Joshua L.}
\surname{Payne}
}

\email{joshua.payne@dartmouth.edu}

\affiliation{
  Computational Genetics Laboratory, 
  Dartmouth College, Hanover, NH 03755
}

\author{
\firstname{Kameron Decker}
\surname{Harris}
}

\email{kameron.harris@uvm.edu}

\affiliation{Department of Mathematics \& Statistics,
  The University of Vermont,
  Burlington, VT 05401.}

\affiliation{Complex Systems Center
  \& the Vermont Advanced Computing Center,
  The University of Vermont,
  Burlington, VT 05401.}

\author{
\firstname{Peter Sheridan}
\surname{Dodds}
}

\email{peter.dodds@uvm.edu}

\affiliation{Department of Mathematics \& Statistics,
  The University of Vermont,
  Burlington, VT 05401.}

\affiliation{Complex Systems Center
  \& the Vermont Advanced Computing Center,
  The University of Vermont,
  Burlington, VT 05401.}

\markboth{Title}
{Author names}

\date{\today}

\begin{abstract}
  
We derive analytic expressions 
for the possibility, probability, and expected size of global spreading events
starting from a single infected seed
for a broad collection of contagion processes acting on random networks
with both directed and undirected edges and arbitrary degree-degree correlations.
Our work extends previous theoretical developments for
the undirected case, and we provide numerical support
for our findings by investigating an example
class of networks for which we are able to obtain closed-form expressions.

\end{abstract}


\maketitle

\section{Introduction}
\label{sec:ccdnw.introduction}

Spreading mechanisms playing out
on generalized random networks
constitute a rich and compelling 
class of tractable contagion 
models~\cite{newman2003a,boccaletti2006a}.
First, while real world complex networks are rarely,
if ever, pure \erdosrenyi\ networks,
they often possess a strong, describable measure of 
randomness~\cite{shen-orr2002a},
once the dominant aspect of
degree distribution is acknowledged~\cite{albert2002a}.
Second, simple models of network-based spreading have
yielded important insights into 
spreading phenomena
such as the spread of infectious diseases~\cite{bajardi2011a,watts2005a},
cascading failures in power grids~\cite{buldyrev2010a,hines2010a},
and
social contagion processes~\cite{schelling1971a,schelling1973a,schelling1978a,granovetter1978a,kuran1991a,kuran1997a,dodds2004a,centola2007a,centola2007b,centola2010a}.
Finally, many random network models
are amenable to analytic investigations
and researchers have naturally built on
areas of statistical mechanics---with its
great tradition of exactly solvable models---such
as the study of percolation on lattices~\cite{stauffer1992a}.

Here, we examine contagion processes acting on mixed directed and undirected 
degree-assortative random networks.
Specifically, for the case of a single seed,
we derive and verify by simulations 
analytic expressions for three key aspects
of these systems:
(1) the \textit{possibility} of a global spreading event; 
(2) the \textit{probability} of a global spreading event;
and
(3) the \textit{expected final size} of a successful global spreading event.
We make the distinction between possibility and probability,
the former referring to the potential for spreading 
(i.e., whether or not the system is in a phase where spreading may occur),
and the latter to the quantified chance that a macroscopic
spreading event may arise given the nature of the initial seed (e.g., random
or targeted).  
To put it another way,
possibility is a categorical yes/no
criterion and probability is a quantitative one;
they ask different kinds of questions, and 
elicit different kinds of analyses for their determination.
Thus, while we could simply derive the probability of global spreading only and thereby 
immediately know if global spreading was possible or not 
(corresponding to non-zero or zero probabilities),
obtaining the possibility of global spreading alone is important as
it directly reveals phase transitions, and further
involves a transparent,
physically argued calculation~\cite{dodds2011b}.

We base our work most strongly on 
two groups of authors' findings:
Bogu\~{n}\'{a} and Serrano~\cite{boguna2005a},
who provided a general formulation for such
networks;
and
Gleeson and Cahalane~\cite{gleeson2007a,gleeson2008a},
who derived the final size of global spreading
events for general contagion models on a wide
array of network structures,
including the social-like threshold model
on random networks~\cite{schelling1971a,granovetter1978a,watts2002a}.
Our work is related to that of Meyers \etal~\cite{meyers2006a}
who examined disease-spreading models on mixed directed and 
undirected uncorrelated networks; 
our analytic methods are essentially disjoint
and we treat more general spreading mechanisms (e.g., social-like ones),
while Meyers \etal\ explored various real-world applications.

We structure our paper as follows.
In Sec.~\ref{sec:ccdnw.modeldescription}, 
we define the family of random networks
and contagion processes we  investigate here;
in Sec.~\ref{sec:ccdnw.contagioncondition}
we provide physically-motivated expressions for 
the possibility and probability
of global spreading events starting
from a single seed;
and
in Sec.~\ref{sec:ccdnw.finalsize}, 
we derive coupled evolution
equations that describe the growth 
of a global spreading event,
as well as yield the expected final size.
In the appropriate limits, our equations 
collapse to those for various network subclasses
involving purely directed or undirected links,
and correlated or uncorrelated nodes.
In Sec.~\ref{sec:ccdnw.toymodel},
we obtain exact results regarding spreading
on a specific family of random networks.
We close with a few remarks in
Sec.~\ref{sec:ccdnw.conclusion}.

\section{Model description}
\label{sec:ccdnw.modeldescription}

\subsection{Generalized random networks}
\label{subsec:ccdnw.networks}

We consider random networks containing
undirected and directed links
along with arbitrary
correlations between nodes based on degrees.
Following Bogu\~{n}\'{a} and Serrano~\cite{boguna2005a},
we allow each node to have 
$\kbid$ undirected edges, 
$\kin$ incoming directed edges,
and 
$\kout$ outgoing directed edges.
We assume all edges are unweighted.
We represent
a node by the generalized degree 
vector $\veck = [\kbid,\kin,\kout]^{\rm T}$,
which we will refer to simply as a node's degree,
and we write $P(\veck)$ for the degree distribution.

To account for correlations,
three conditional probabilities are
needed:
$\Probbid(\veck\, |\, \veck')$, 
$\Probin(\veck\, |\, \veck')$, 
and
$\Probout(\veck\, |\, \veck')$;
these quantities give the 
chances that an edge starting
at a degree $\veck'$ node
ends at a degree $\veck$ node 
and is respectively undirected,
incoming, or outgoing 
relative to
the destination degree $\veck$ node
(note that this convention for directed
edges is opposite that used in~\cite{boguna2005a}).

For networks to be well defined (i.e., realizable),
these probabilities must be constrained 
by two detailed balance equations.
In determining the probability that an edge of a certain
type runs between nodes of degree $\veck$ and $\veck'$,
we must obtain the same result whether we start at
the former or latter node.
We write
the probability that a randomly chosen edge is
undirected and connects a degree $\veck$
and degree $\veck'$ node as $\Probbid(\veck, \veck')$.
Noting that 
the probability that a random end
of a randomly selected undirected edge emanates
from a degree $\veck$ node is given by
$\frac{\kbid P(\veck)}{\tavg{\kbid}}$,
we first have that
\begin{eqnarray}
  \Probbid(\veck, \veck')
  & = &
  \Probbid(\veck\, |\, \veck') 
  \frac{\kbid' P(\veck')}{\tavg{\kbid'}}
  \nonumber \\ 
  & = &
  \Probbid(\veck'\, |\, \veck) 
  \frac{\kbid P(\veck)}{\tavg{\kbid}}
  =  \Probbid(\veck', \veck).
  \label{eq:ccdnw.detailedbalance1}
\end{eqnarray}
For directed edges, we define
$\Probdir(\veck,\veck')$ as the
probability that a randomly chosen
edge is directed and leads from a degree $\veck'$
node to a degree $\veck$ node.
Similar to the balance equation for
undirected edges, 
we use the quantities
$\frac{\kout P(\veck)}{\tavg{\kout}}$
and
$\frac{\kin P(\veck)}{\tavg{\kin}}$
which give the probabilities
that in starting at a random end
of a randomly selected edge,
we begin at a degree $\veck$ node
and then
find ourselves travelling
(1) along an outgoing edge
or 
(2) against the direction of an incoming edge.
We therefore have
\begin{equation}
  \Probdir(\veck,\veck')
  =
  \Probin(\veck\, |\, \veck') 
  \frac{\kout' P(\veck')}{\tavg{\kout'}}
  =
  \Probout(\veck'\, |\, \veck) 
  \frac{\kin P(\veck)}{\tavg{\kin}}.
  \label{eq:ccdnw.detailedbalance2}
\end{equation}
Note that since 
$\tavg{\kbid}=\tavg{\kbid'}$
and
$\tavg{\kout'}=\tavg{\kout}=\tavg{\kin}$,
the denominators in 
Eqs.~\req{eq:ccdnw.detailedbalance1}
and~\req{eq:ccdnw.detailedbalance2}
are equal and may be omitted~\cite{boguna2005a}.
Further, our alternate definitions
of $\Probin(\veck'\, |\, \veck)$
and $\Probout(\veck'\, |\, \veck)$
mean that 
\Req{eq:ccdnw.detailedbalance2}
has a form different to that given in~\cite{boguna2005a}.

For the class of random networks given above,
Bogu\~{n}\'{a} and Serrano
determine a number of structural results
regarding percolation,
including the sizes of the giant in-component,
out-component, and strongly connected component~\cite{boguna2005a}.
Our goal here is to examine the behavior of generalized
spreading processes on such networks,
and we describe these next.

\subsection{Contagion processes}
\label{subsec:ccdnw.contagionprocesses}

We consider synchronous discrete time contagion processes,
though our results can at least in part 
be extended to asynchronous discrete
and continuous time processes~\cite{gleeson2008a}.
We assume that once nodes are infected, they
remain so permanently, an aspect that is needed
for computing the final size of a global spreading event.
We write the probability of node $j$ becoming infected
in time step $t+1$ as
\begin{equation}
  \label{eq:ccdnw.probinf}
  \indrspfn_{j}
  (
  k_{\rm inf};
  \kbid + \kin
  ),
\end{equation}
given that $k_{\rm inf}$ of
node $j$'s 
total of $\kbid + \kin$
undirected and incoming edges 
emanate from infected nodes at time $t$.
Here, $\indrspfn_{j}$ is an arbitrary, node-specific `response function'
mapping to the unit interval.
Now, for the general class of contagion models we consider here on 
infinite random networks, 
we need to know only the average response
function for each node subclass.
Taking all nodes of degree $\veck$,
having indices in the set 
$J_{\veck} = \{ j_{\veck,1}, j_{\veck,2}, \ldots, j_{\veck,n}, \ldots\}$,
we compute this average response function as 
\begin{equation}
  \label{eq:ccdnw.avgresponsefn}
  \rspfn_{k_{\rm inf},\veck}
  =
  \lim_{n \rightarrow \infty}
  \frac{1}{n}
  \sum_{j = j_{\veck,1}}^{j_{\veck,n}}
  \indrspfn_{j}
    (
  k_{\rm inf};
  \kbid + \kin 
  ).
\end{equation}
The quantity $\rspfn_{k_{\rm inf},\veck}$
is then the probability that a randomly chosen
node of degree $\veck$ is infected at time $t+1$
given that at time $t$, it has
$k_{\rm inf}$ infected incoming and undirected
edges.

\section{Possibility and probability of global spreading}
\label{sec:ccdnw.contagioncondition}

In~\cite{dodds2011b}, we derived a global spreading condition
for discrete and continuous time contagion processes
with the possibility of recovery
acting on generalized random networks.
Defining $\vec{\alpha}=(\nu,\lambda)$
to represent a pairing of a type $\nu$ node
and type $\lambda$ edge, we argued
that the number of infected node-edge
pairs $f_{\vec{\alpha}}$
grows as a function of network distance
$d$ from a seed as
$
  f_{\vec{\alpha}}(d+1) 
  = 
  \sum_{\vec{\alpha}'}
  R_{\vec{\alpha} \vec{\alpha}'}
  f_{\vec{\alpha}'}(d),
$
where $R_{\vec{\alpha} \vec{\alpha}'}$ 
depends simply on network structure
and the spreading process~\cite{dodds2011b}.
As a special but still broad case,
we showed that for the networks we consider here,
the growth rate equation for the number of 
infected edges emanating from degree $\veck$ nodes
a distance $d$ from an initiating node 
obeys the following:
\begin{equation}
  \left[
    \begin{array}{c}
      f_{\veck}^{\rm (\bidmark)}(d+1) \\
      f_{\veck}^{\rm (\outmark)}(d+1)
    \end{array}
  \right]
  =
  \sum_{k'}
  \textbf{R}_{\veck\veck'}
  \left[
    \begin{array}{c}
      f_{\veck'}^{\rm (\bidmark)}(d) \\
      f_{\veck'}^{\rm (\outmark)}(d)
    \end{array}
  \right],
  \label{eq:ccdnw.R}
\end{equation}
where
\begin{eqnarray}
  \lefteqn{
    \textbf{R}_{\veck\veck'}
    = 
  } \\
  & &
    \left[
    \begin{array}{cc}
      \Probbid(\veck\,|\, \veck')
      \bullet
      (\kbid-1)
      &
      \Probin(\veck\,|\, \veck')
      \bullet
      \kbid
      \\
      \Probbid(\veck\,|\, \veck')
      \bullet
      \kout
      &
      \Probin(\veck\,|\, \veck')
      \bullet
      \kout
    \end{array}
    \right]
    \bullet
    \rspfn_{1,\veck}.
    \nonumber
  \label{eq:ccdnw.Rkkdef}
\end{eqnarray}
Here, the quantities 
$f_{\veck}^{\rm (\bidmark)}(d)$
and 
$f_{\veck}^{\rm (\outmark)}(d)$
are the number of `infected' undirected and outgoing 
edges leaving an infected degree $\veck$ node a distance
$d$ steps from the seed.
We have expressed the form of $\textbf{R}_{\veck\veck'}$
so as to make clear the three components making
up general spreading conditions:
(1) probability of connection
[$\Probbid(\veck\,|\, \veck')$ 
and
$\Probin(\veck\,|\, \veck')$];
(2) resultant newly infected edges
[$(\kbid-1)$, $\kbid$ and $\kout$ factors];
and 
(3) the probability of infection
($\rspfn_{1,\veck}$)~\cite{dodds2011b}.
The above agrees with the contagion condition found earlier by Bogu\~n\'a
and Serrano for the emergence of the giant out-component using
a generating function approach.
Note that these calculations depend on the local pure branching
structure of random networks with zero clustering;
for recent advances for the non-zero clustering
case, see~\cite{gleeson2010a,ikeda2010a,hackett2011a,melnik2011a}.

The full gain matrix $\mathbf{R}$ and edge infection counts $\vec{f}(d)$
can be laid out as follows:
\begin{equation}
  \mathbf{R} =
  \left[
  \begin{array}{ccc}
  \mathbf{R}_{\veck_1\veck_1} & \mathbf{R}_{\veck_1\veck_2} & \ldots \\
  \mathbf{R}_{\veck_2\veck_1} & \mathbf{R}_{\veck_2\veck_2} & \ldots \\
  \vdots & \vdots & \ddots \\
  \end{array}
  \right]
  \mbox{\ and\ }
  \vec{f}(d) =
  \left[
  \begin{array}{c}
    f_{\veck_1}^{\rm (\bidmark)}(d) \\
    f_{\veck_1}^{\rm (\outmark)}(d) \\
    f_{\veck_2}^{\rm (\bidmark)}(d) \\
    f_{\veck_2}^{\rm (\outmark)}(d) \\
    \vdots \\
  \end{array}
  \right].
  \label{eq:ccdn.Rmatrix}
\end{equation}

The condition for the possibility of global spreading events
is therefore that the maximum eigenvalue of 
$
\left[
  \textbf{R}_{\veck\veck'}
\right]
$
exceeds 1:
\begin{equation}
  \sup
  \left\{
    |\mu| : \mu \in 
    \sigma
    \left(
      \left[
        \textbf{R}_{\veck\veck'}
      \right]
    \right)
  \right\}
  >
  1
    \label{eq:ccdnw.maxeig}
\end{equation}
where $\sigma(\cdot)$ indicates spectrum.

Next, we determine the probability of a global spreading
event given the initial seed is of degree $\veck$
and hence the overall probability given a randomly
selected seed; 
we refer to these quantities as
`triggering' probabilities.
While in determining the
probability of a global spreading event we must
also determine the possibility, the direct calculation
we have just presented for the latter is needed
to demonstrate a physically-motivated clarity.

We define
$\probfc^{(\rm u)}_{\veck}$
to be the probability
that an infected undirected
edge leaving a degree $\veck$ node 
will lead to a giant component of
infected nodes.
Similarly, we 
define 
$\probfc^{(\rm o)}_{\veck}$
to be the probability that
an infected outgoing edge from a degree $\veck$ node
will generate a global spreading event.
Using the Markov nature of 
random networks, we can write
down recursive, closed-form relationships
for these two probabilities:
\begin{eqnarray}
  \label{eq:ccdnw.Qu}
  \lefteqn{\probfc^{(\rm u)}_{\veck} = }  \\
  & & 
  \sum_{\veck'}
  \Probbid(\veck' | \, \veck)
  \left[
    1
    -
    \left(1-\probfc^{(\rm u)}_{\veck'}\right)^{\kbid'-1}
    \left(1-\probfc^{(\rm o)}_{\veck'}\right)^{\kout'}
  \right]
  \rspfn_{1,\veck'}, \nonumber
\end{eqnarray}
and
\begin{eqnarray}
  \label{eq:ccdnw.Qo}
  \lefteqn{\probfc^{(\rm o)}_{\veck} = } \\
  & &
  \sum_{\veck'}
  \Probin(\veck' | \, \veck)
  \left[
    1
    -
    \left(1-\probfc^{(\rm u)}_{\veck'}\right)^{\kbid'}
    \left(1-\probfc^{(\rm o)}_{\veck'}\right)^{\kout'}
  \right]
  \rspfn_{1,\veck'}. 
  \nonumber
\end{eqnarray}
In these equations, we have encoded the understanding
that if an infected edge generates a global spreading event,
then it must infect its target node which in turn must
be successful in infecting its other neighbors.
In~\Req{eq:ccdnw.Qo}, for example,
$\Probin(\veck' | \, \veck)$ is the probability
that the undirected edge leads from
an infected degree $\veck$ node 
to a degree $\veck'$ node
which it infects with probability
$\rspfn_{1,\veck'}$.
The quantity
$
(1-\probfc^{(\rm u)}_{\veck'})^{\kbid'}
(1-\probfc^{(\rm o)}_{\veck'})^{\kout'}
$
is the probability that none of the infected node's
other undirected or outgoing edges successfully spread
the infection, and hence 
$
  \left[
    1
    -
    (1-\probfc^{(\rm u)}_{\veck'})^{\kbid'}
    (1-\probfc^{(\rm o)}_{\veck'})^{\kout'}
  \right]
$
is the probability that at least one does.

Both 
$\probfc^{(\rm u)}_{\veck}$
and
$\probfc^{(\rm o)}_{\veck}$
can be determined from
Eqs.~\req{eq:ccdnw.Qu}
and~\req{eq:ccdnw.Qo}
either numerically or 
exactly 
(as per our example later in Sec~\ref{sec:ccdnw.toymodel}).
Having done so, we can then compute the probability
that infecting a single degree $\veck$ node triggers a global spreading event:
\begin{equation}
  \label{eq:ccdnw.ptrigk}
  \Ptrig(\veck)
  =
  \left[
    1 - (1-\probfc^{\rm (u)}_{\veck})^{\kbid}
    (1-\probfc^{\rm (o)}_{\veck})^{\kout}
  \right],
\end{equation}
which is the complement of 
$(1-\probfc^{\rm (u)}_{\veck})^{\kbid}(1-\probfc^{\rm (o)}_{\veck})^{\kout}$,
the probability of failure to trigger.
The probability that infecting a randomly chosen
node triggers a global spreading event 
is then simply $\Ptrig = \sum_{\veck} \Ptrig(\veck)$, or 
\begin{equation}
  \label{eq:ccdnw.ptrigrandom}
  \Ptrig
  =
  \sum_{\veck} 
  P(\veck)
  \left[
    1 - (1-\probfc^{\rm (u)}_{\veck})^{\kbid}
    (1-\probfc^{\rm (o)}_{\veck})^{\kout}
  \right].
\end{equation}
In similar fashion, the triggering probability for 
non-random, strategic selections of the initial seed can
readily be obtained.  
Appropriate limits of~\Req{eq:ccdnw.ptrigrandom}
also recover triggering probabilities for
simpler families of random networks such
as undirected, uncorrelated networks with
prescribed degree distributions.
Finally, considering the limit of $\Ptrig \rightarrow 0$
retrieves the condition for global spreading
found above.

\section{Final size of successful global spreading events}
\label{sec:ccdnw.finalsize}

We complete our main analysis by determining
the final size of a global spreading event
building on the work of Gleeson and Cahalane~\cite{gleeson2007a}
and later Gleeson~\cite{gleeson2008a}.
We shift our focus from spreading away from a seed (expansion)
to spreading reaching a node (contraction).

We consider an arbitrary fixed node in the network
and compute the probability that incoming edges (directed
or undirected) are infected and sufficient in number
that the node itself becomes infected at a certain time.
To do so, we need to first determine the probabilities
that undirected and incoming edges arriving 
at a degree $\veck$ node are infected at time $t$,
$\theta_{\veck,t}^{(\bidmark)}$ and
$\theta_{\veck,t}^{(\inmark)}$.
As with the possibility and probability of spreading,
edge-edge transitions are the best framing
for this calculation.  Edges will be infected at time
$t+1$ if the node from which they emanate becomes
infected in that time step, and this in turn depends
on the infection levels of the incoming edges.
Assuming a fraction $\phi_0>0$ of initially infected
seeds in the network, we obtain the following expression
for the fraction of infected directed and incoming 
edges in the network at time $t+1$:
\begin{multline}
  \theta_{\veck,t+1}^{(\bidmark)} = 
  \phi_0 \\
  + 
  (1-\phi_0)
  \sum_{\veck'} 
  P^{(\bidmark)} ( \edgetype )
  \sum_{\jbid=0}^{\kbid'-1}
  \sum_{\jin=0}^{\kin'} 
  \binom{\kbid'-1}{\jbid} 
  \binom{\kin'}{\jin}
  \\
  \times
  \left[
    \theta_{\veck',t}^{(\bidmark)}
  \right]^{\jbid}
  \left[
    1-\theta_{\veck',t}^{(\bidmark)}
  \right]^{(\kbid'-1-\jbid)} 
  \left[
    \theta_{\veck',t}^{(\inmark)}
  \right]^{\jin}
  \left[
    1-\theta_{\veck',t}^{(\inmark)}
  \right]^{(\kin'-\jin)} 
  \\
  \times
  \rspfn_{\jbid+\jin,\kbid'+\kin'},
  \label{eq:ccdnw.edgeEvU}
\end{multline}
and
\begin{multline}
  \theta_{\veck,t+1}^{(\inmark)} = 
  \phi_0 \\
  + 
  (1-\phi_0)
  \sum_{\veck'} 
  P^{(\inmark)} ( \edgetype )
  \sum_{\jbid=0}^{\kbid'} 
  \sum_{\jin=0}^{\kin'} 
  \binom{\kbid'}{\jbid} 
  \binom{\kin'}{\jin}
  \\ \times
  \left[
    \theta_{\veck',t}^{(\bidmark)}
  \right]^{\jbid}
  \left[
    1-\theta_{\veck',t}^{(\bidmark)}
  \right]^{(\kbid'-\jbid)} 
  \left[
    \theta_{\veck',t}^{(\inmark)}
  \right]^{\jin}
  \left[
    1-\theta_{\veck',t}^{(\inmark)}
  \right]^{(\kin'-\jin)} 
  \\
  \times
  \rspfn_{\jbid+\jin,\kbid'+\kin'}.
  \label{eq:ccdnw.edgeEvI}
\end{multline}
Since we are now considering contraction rather than
expansion, more than one edge may contribute to
the infection of a node, hence the sum over
nearly the full range of infection probabilities,
the $\{\rspfn_{\jbid+\jin,\kbid'+\kin'}\}$.

The overall fraction of infected nodes at time $t$,
equivalently the probability that a randomly
chosen node becomes infected at time $t$, 
depends on $\theta_{\veck',t}^{(\bidmark)}$
and $\theta_{\veck',t}^{(\inmark)}$ as
\begin{multline}
  \phi_{t+1} = \phi_0 \\
  + 
  (1-\phi_0)
  \sum_{\veck} 
  P(\veck)
  \sum_{\jbid=0}^{\kbid}
  \sum_{\jin=0}^{\kin} 
  \binom{\kbid}{\jbid}
  \binom{\kin}{\jin}
  \\
  \times
  \left[
    \theta_{\veck,t}^{(\bidmark)}
  \right]^{\jbid}
  \left[
    1-\theta_{\veck,t}^{(\bidmark)}
  \right]^{(\kbid-\jbid)} 
  \left[
    \theta_{\veck,t}^{(\inmark)}
  \right]^{\jin}
  \left[
    1-\theta_{\veck,t}^{(\inmark)}
  \right]^{(\kin-\jin)} 
  \\
  \times
  \rspfn_{\jbid+\jin,\kbid+\kin}.
  \label{eq:ccdnw.phifinalsize}
\end{multline}
To determine the final size, we set 
$
\theta_{\veck',t+1}^{(\bidmark)}
=
\theta_{\veck',t}^{(\bidmark)}
=
\theta_{\veck',\infty}^{(\bidmark)}
$
and
$
\theta_{\veck',t+1}^{(\inmark)}
=
\theta_{\veck',t}^{(\inmark)}
=
\theta_{\veck',\infty}^{(\inmark)}
$
in Eqs.~\req{eq:ccdnw.edgeEvU}
and
\req{eq:ccdnw.edgeEvI}
and solve for the steady-state solutions
$\theta_{\veck',\infty}^{(\bidmark)}$
and
$\theta_{\veck',\infty}^{(\inmark)}$.
Substituting these values
into~\Req{eq:ccdnw.phifinalsize}
gives us the expected final size
$\phi_{\infty}$ which is, among other things,
a function of $\phi_0$, the fraction of nodes
initially infected.  For the single seed
case we consider in this present work,
the final step therefore is 
to take the limit $\phi_0 \rightarrow 0$.
Note that as for the triggering probability,
the condition for global spreading,
\Req{eq:ccdnw.maxeig}, can be recovered
by linearizing
Eqs.~\req{eq:ccdnw.edgeEvU},
\req{eq:ccdnw.edgeEvI},
and~\req{eq:ccdnw.phifinalsize} (see ref.~\cite{gleeson2008a}).

\section{Exact solution for an example degree-correlated random network with mixed directed and undirected edges}
\label{sec:ccdnw.toymodel}

\begin{figure*}[tp!]
  \centering
  \includegraphics[width=\textwidth]{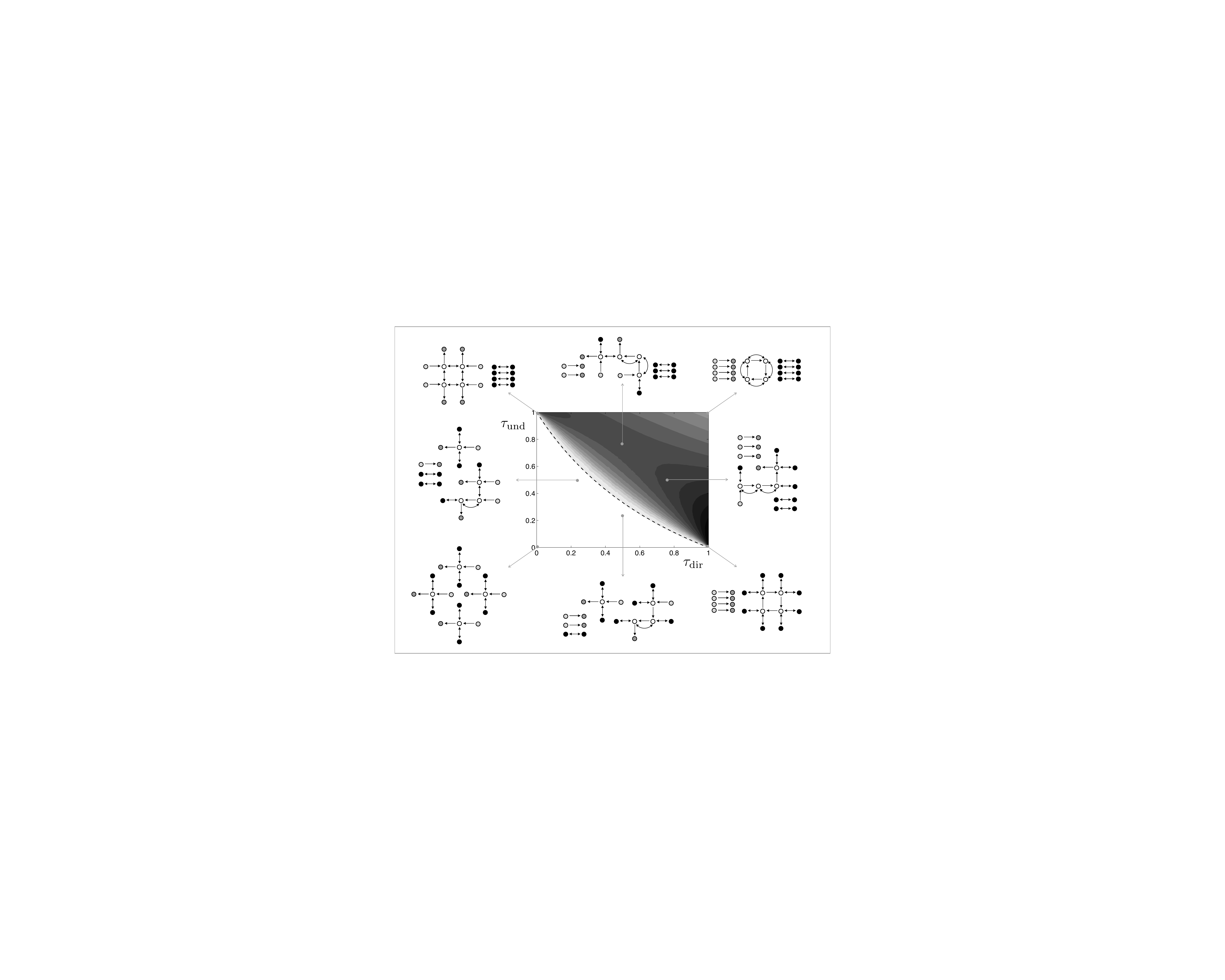}
  \caption{
    Central plot: For the toy network model described in Sec.~\ref{sec:ccdnw.toymodel},
    final size $\phi_{\infty}$ as 
    a function of the model parameters $\toyund$ and $\toydir$.
    Size is mapped to a linear gray scale with white indicating no global spreading.
    The dashed line marks the theoretically determined phase transition 
    given in~\Req{eq:ccdnw.mulambdacurve}.
    The example networks shown around the plot give a sense
    of the kinds of networks realized for the corresponding $\toyund$ and $\toydir$.
    Note that we show exact forms for networks with 20 nodes to make clear
    how node types are correlated in phase space, and forms will
    differ for larger networks.
    For example, for $\toyund=\toydir=1$, networks will comprise
    a giant component of type 1 nodes (open circles) with the remaining
    three node types represented only in isolated pairs.
    Simulation details:
    We formed each network with $N = 10^4$ nodes made up of a 1:1:1:2 ratio of node types
    1 through 4.
    We constructed each network initially to have $\toyund = \toydir = 1$,
    which was simple algorithmically,
    and then shuffled edges until desired values of $\toyund$ and $\toydir$ were reached,
    using an approach similar to those described in~\cite{milo2003a} and~\cite{dodds2009a}.
    We further shuffled each edge type 10,000 times to ensure randomization.
    For each $\toyund$ and $\toydir$ in 0, 0.01, 0.02, \ldots, 1.00, we generated
    100 networks and randomly picked 1000 seeds for a total of $10^5$ samples.
  }
  \label{fig:ccdnw.schematic}
\end{figure*}

To test our analytic expressions for the 
possibility, probability, and expected final size of 
a global spreading event,
we consider a family of general random networks
for which our equations are exactly solvable.
As shown schematically in the margins 
of Fig.~\ref{fig:ccdnw.schematic}, we allow four
types of nodes with the following degree vectors,
which, again, have the form
$\veck = [\kbid, \kin, \kout]^{\rm T}$:
\begin{equation}
  \label{eq:ccdnw.toydegreevecs}
  \veck_{1} = 
  \left[
    \begin{array}{c}
      2 \\
      1 \\
      1 \\
    \end{array}
  \right],
  \veck_{2} = 
  \left[
    \begin{array}{c}
      0 \\
      0 \\
      1 \\
    \end{array}
  \right],
  \veck_{3} = 
  \left[
    \begin{array}{c}
      0 \\
      1 \\
      0 \\
    \end{array}
  \right],
  \ \mbox{and} \
  \veck_{4} = 
  \left[
    \begin{array}{c}
      1 \\
      0 \\
      0 \\
    \end{array}
  \right],
\end{equation}
and which occur with abundances
\begin{gather}
   \label{eq:ccdnw.toydegreeprobs} 
   P(\veck_{1}) = \frac{1}{5},
   P(\veck_{2}) = \frac{1}{5}, 
   P(\veck_{3}) = \frac{1}{5},
   \ \mbox{and} \
   P(\veck_{4}) = \frac{2}{5}.
\end{gather}
We define the degree-degree conditional probabilities 
as dependent on two tunable parameters,
$\toyund$ and $\toydir$:
\begin{equation}
  \left[
  P^{(\bidmark)}(\edgetype) 
  \right]
  = 
  \begin{bmatrix}
    \toyund & 0 & 0 & (1-\toyund) \\
    0 & 0 & 0 & 0 \\
    0 & 0 & 0 & 0 \\
    (1-\toyund) & 0 & 0 & \toyund 
  \end{bmatrix},
\end{equation}
and
\begin{gather}
  \left[
    P^{(\inmark)}(\edgetype)
  \right]
  =
  \left[
    P^{(\outmark)}(\edgetype)
  \right]^{\rm T}
  =
  \nonumber \\
  \begin{bmatrix}
    \toydir & (1-\toydir) & 0 & 0 \\
    0 & 0 & 0 & 0  \\
    (1-\toydir) & \toydir & 0 & 0 \\
    0 & 0 & 0 & 0  \\
  \end{bmatrix}.
\end{gather}
where $0 \le \toydir, \toyund \le 1$,
and $\veck$ and $\veck'$ correspond
to rows and columns.

We have chosen
$\toydir$ and $\toyund$ so that increasing
them will tend to increase global connectivity,
with $\toyund$ controlling correlations
between nodes through undirected edges,
and $\toydir$ through directed ones.
There are four clear limiting cases,
as shown in the corners of Fig.~\ref{fig:ccdnw.schematic}.
For example, when $\toydir=\toyund=1$
(upper right corner of Fig.~\ref{fig:ccdnw.schematic}), 
type 1 nodes
are connected only to other type 1 nodes
creating a giant component, while the other
three types combine to form isolated
pairs with either directed
or undirected connections. 
At the other extreme when $\toydir=\toyund=0$, 
(lower left corner of Fig.~\ref{fig:ccdnw.schematic}), 
each of the
four edges from type 1 nodes
connect only to type 2, 3, and 4 nodes,
meaning the network is composed
of discrete, five-node components.
The six other example networks
in Fig.~\ref{fig:ccdnw.schematic}
give a sense of the other possible 
configurations contained within this
simple network family we have constructed.

We obtain results for general response functions,
while for comparison with simulations, 
we consider a test contagion process
with the following single-parameter threshold transmission probabilities:
\begin{gather}
  \rspfn_{0,\veck_i}=0,
  \rspfn_{1,\veck_1}=\infparam,
  \mbox{\ and\ }
  \rspfn_{j,\veck_i}=1
  \mbox{\ otherwise.}
\label{eq:ccdn.toyresponse}
\end{gather}
where $i = 1, \ldots, 4$.
The choice $\rspfn_{0,\veck_i}=0$ means
no nodes spontaneously become infected 
(as might model the action of an exogenous source of infection).
In the case that $\infparam=1$, then
this set of responses means that if a node finds at least one 
neighbor at the end
of an undirected or incoming edge that is infected,
then the node itself becomes infected in the next time step.
For $\infparam<1$, a random fraction $\infparam$ of degree $\veck_1$ nodes 
become infected in the time step following the infection
of a single neighbor whereas $1-\infparam$ remain uninfected.
As discussed in Sec.~\ref{subsec:ccdnw.contagionprocesses},
individual response functions need only give this
average response function; for example, a fraction
$\infparam$ of degree $\veck_1$ nodes might
have a deterministic threshold of 1 with 
the remaining fraction of $1-\infparam$ having a 
deterministic threshold of 2.

Returning to Fig.~\ref{fig:ccdnw.schematic},
the gray-scale plot shows the fractional size of
successful global spreading events as a function of
$\toyund$ and $\toydir$ for the specific
spreading mechanism described above
with $\infparam (=\rspfn_{1,\veck_1})=1$.
We see a clear phase transition indicated
by the dashed curve and our next task is
to find its analytic form.

\subsection{Global spreading condition}
\label{subsec:ccdnw.toymodel-possibility}

Using the spreading conditions
contained in Eqs.~\req{eq:ccdnw.R}, \req{eq:ccdnw.Rkkdef}, 
and \req{eq:ccdnw.maxeig},
and the model's definition,
we find that global spreading may occur
when the maximum eigenvalue of the following gain matrix 
exceeds unity:
\begin{widetext}
\begin{equation}
  \label{eq:ccdnw.toymatrix}
  \mathbf{R} =
  \left[
    \begin{array}{cccc}
      \mathbf{R}_{\veck_1\veck_1} & \mathbf{R}_{\veck_1\veck_2} & \mathbf{R}_{\veck_1\veck_3} & \mathbf{R}_{\veck_1\veck_4} \\
      \mathbf{R}_{\veck_2\veck_1} & \mathbf{R}_{\veck_2\veck_2} & \mathbf{R}_{\veck_2\veck_3} & \mathbf{R}_{\veck_2\veck_4} \\
      \mathbf{R}_{\veck_3\veck_1} & \mathbf{R}_{\veck_3\veck_2} & \mathbf{R}_{\veck_3\veck_3} & \mathbf{R}_{\veck_3\veck_4} \\
      \mathbf{R}_{\veck_4\veck_1} & \mathbf{R}_{\veck_4\veck_2} & \mathbf{R}_{\veck_4\veck_3} & \mathbf{R}_{\veck_4\veck_4} \\
    \end{array}
  \right]
  =
  \left[
    \begin{array}{cccccccc}
      \toyund \rspfn_{1,\veck_1} 
      & 
      2\toydir \rspfn_{1,\veck_1} 
      & 
      0 
      & 
      2(1-\toydir) \rspfn_{1,\veck_2} 
      &
      0 
      & 
      0 
      & 
      (1-\toyund) \rspfn_{1,\veck_4} 
      & 
      0 
      \\
      \toyund \rspfn_{1,\veck_1}
      & 
      \toydir \rspfn_{1,\veck_1}
      & 
      0 
      & 
      (1-\toydir) \rspfn_{1,\veck_2} 
      & 
      0 
      & 
      0 
      & 
      (1-\toyund) \rspfn_{1,\veck_4}
      & 
      0 
      \\
                  0 & 0 & 0 & 0 & 0 & 0 & 0 & 0 \\
      \vdots & \vdots & \vdots & \vdots & \vdots & \vdots & \vdots & \vdots \\ 
      0 & 0 & 0 & 0 & 0 & 0 & 0 & 0 \\
    \end{array}
  \right].
\end{equation}
\end{widetext}
Clearly, only the top left hand corner of this 
gain matrix matters as global spreading, if possible,
must occur on a giant component.
Upon substitution of the model's response functions
given in~\Req{eq:ccdn.toyresponse},
we find the largest eigenvalue is 
$
\frac{1}{2}
(\toyund+\toydir+\sqrt{(\toyund+\toydir)^2 + 4\toyund\toydir})
\infparam
$
and we find that global spreading events therefore occur in the region
described by 
\begin{equation}
  \label{eq:ccdnw.mulambdacurve}
  (1+\toyund \infparam)(1+\toydir \infparam) > 2.
\end{equation}
For the case $\infparam (=\rspfn_{1,\veck_1})=1$,
this equation is indeed represented by the dashed curve
shown in Fig.~\ref{fig:ccdnw.schematic},
perfectly matching the phase transition
demonstrated by our simulations.
We can also now readily determine
that spreading may occur for
some values of $\toyund$ and $\toydir$
providing $\infparam > \sqrt{2}-1$.

\subsection{Probability of global spreading}
\label{subsec:ccdnw.toymodel-probability}

In computing the probability that a degree $\veck$
node initiates a global spreading event, we observe
that because only type 1 nodes can transmit an
infection, we need only solve the recursion
equations given
in~\req{eq:ccdnw.Qu} and~\req{eq:ccdnw.Qo}
for 
$
\probfc^{(\rm u)}_{\veck_1}
$
and
$
\probfc^{(\rm o)}_{\veck_1}
$.
Nodes of type 2 and 4,
possessing one outgoing and one undirected
edge respectively, may trigger 
global spreading but obviously cannot be 
involved in transmission, and nodes of type 3
can neither start nor help spread an outbreak.
Eqs.~\req{eq:ccdnw.Qu} and~\req{eq:ccdnw.Qo}
reduce to the nonlinear coupled equations:
\begin{equation}
  \probfc^{(\rm u)}_{\veck_1} = 
  \toyund
  \left[
    1
    -
    \left(1-\probfc^{(\rm u)}_{\veck_1}\right)
    \left(1-\probfc^{(\rm o)}_{\veck_1}\right)
  \right]
  \infparam,
  \label{eq:ccdnw.Qutoy}
\end{equation}
and
\begin{equation}
  \probfc^{(\rm o)}_{\veck_1} = 
  \toydir
  \left[
    1
    -
    \left(1-\probfc^{(\rm u)}_{\veck_1}\right)^{2}
    \left(1-\probfc^{(\rm o)}_{\veck_1}\right)
  \right]
  \infparam.
  \label{eq:ccdnw.Qotoy}
\end{equation}
The equations are solvable and we find
\begin{eqnarray}
  \lefteqn{
  \Quk = 
  1
  +
  \frac{1}{2}
  \toyund
  \infparam
  }
  \nonumber \\
  &
  &
  -
  \sqrt{
    \frac{1}{4}
    \left(
      \toyund
      \infparam
    \right)^2
    -
    \frac{
      \toyund
    }
    {
      \toydir
    }
    +
    \frac{
      1
    }
    {
      \toydir
      \infparam
    }
  },
  \label{eq:ccdnw.Qutoysoln}
\end{eqnarray}
with
\begin{equation}
  \Qok 
  =
  \left(
    \frac{1}{\toyund\beta} - 1
  \right)
  \frac{
    \Quk
  }
  {
    1 - \Quk
  }.
  \label{eq:ccdnw.Qotoysoln}
\end{equation}
For $\beta=\toyund=\toydir=1$, 
$\Quk = \Qok = 1$.
In turn,
$\probfc^{(\rm o)}_{\veck_2}$
and
$\probfc^{(\rm u)}_{\veck_4}$
can be expressed in terms of
$\Quk$ and $\Qok$:
\begin{equation}
  \probfc^{(\rm o)}_{\veck_2}
  =
  (1-\toydir)
  \left[
    1
    -
    \left(1-\probfc^{(\rm u)}_{\veck_1}\right)^2
    \left(1-\probfc^{(\rm o)}_{\veck_1}\right)
  \right]
  \infparam,
  \label{eq:ccdnw.toyQo2}
\end{equation}
and
\begin{equation}
  \probfc^{(\rm u)}_{\veck_4}
  =
  (1-\toyund)
  \left[
    1
    -
    \left(1-\probfc^{(\rm u)}_{\veck_1}\right)
    \left(1-\probfc^{(\rm o)}_{\veck_1}\right)
  \right]
  \infparam.
  \label{eq:ccdnw.toyQu4}
\end{equation}

A first check on these
triggering probability expressions 
is that they are in agreement with
the phase transition recorded in
\Req{eq:ccdnw.mulambdacurve}; 
in other words, $\Quk$ and $\Qok$
should vanish along the phase transition.
We see that upon setting 
the right hand side of~\Req{eq:ccdnw.Qutoysoln}
to zero, rearrangement indeed leads to
the condition 
$(1+\toyund \infparam)(1+\toydir \infparam) = 2$.

\begin{figure}[tp!]
  \centering
  \includegraphics[width=.49\textwidth]{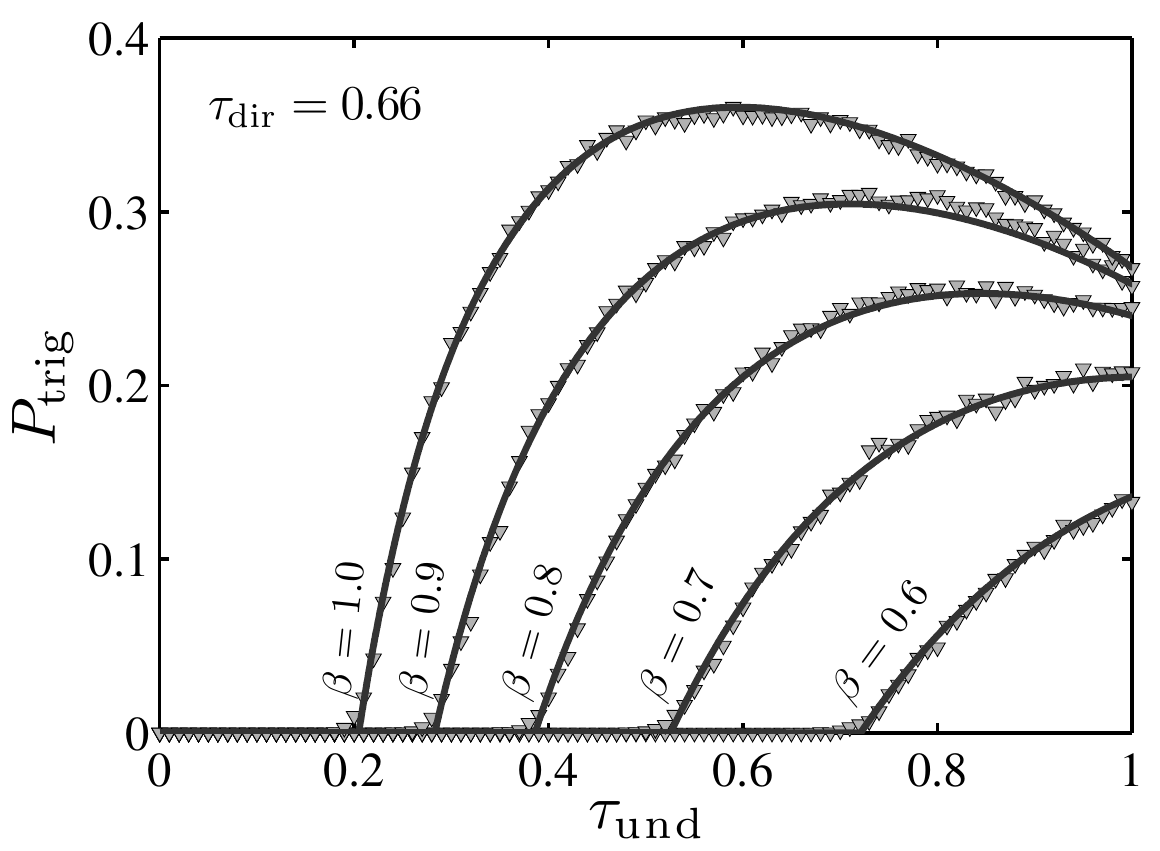}
  \caption{
    For the model described in Sec.~\ref{sec:ccdnw.toymodel},
    the probability that infecting a randomly chosen
    node leads to a global
    spreading event, $\Ptrig$, as a function
    of undirected edge assortativity 
    $\toyund$,
    a fixed value of
    directed edge assortativity, $\toydir=0.66$,
    and varying values of $\infparam =\rspfn_{1,\veck_1}$.
    The curves correspond to output from
    simulations [squares]
    and theory 
    [solid line, \Req{eq:ccdnw.toyptrigk}].
    For the simulation results, we recorded a successful
    global spreading event if the final size exceeded 2.5\% of the network.
    This cut off is arbitrary but nearby values do not
    appreciably change the resulting picture because
    above the phase transition the final size is bimodal:
    either spreading takes off and reaches a characteristic fraction
    of the network, or it fails.
    The network size is $N=10^5$ and the resolution in $\toyund$ is 0.01.
    See caption of Fig.~\ref{fig:ccdnw.schematic} for further details.
  }
  \label{fig:ccdnw.musliceQu}
\end{figure}

We compute the triggering probability
given a randomly chosen seed using~\Req{eq:ccdnw.ptrigrandom}:
\begin{eqnarray}
  \Ptrig & = & 
  \frac{1}{5}
  \left[
    1 - 
    \left(1-\probfc^{\rm (u)}_{\veck_1}\right)^{2}
    \left(1-\probfc^{\rm (o)}_{\veck_1}\right)
  \right]
  \nonumber \\
  & & 
  +
  \frac{1}{5} 
  \probfc^{\rm (o)}_{\veck_2}
  +
  \frac{2}{5}
  \probfc^{\rm (u)}_{\veck_4}.
  \label{eq:ccdnw.toyptrigk}
\end{eqnarray}
We compare our theoretical
computation of $\Ptrig$ 
with simulations in
Fig.~\ref{fig:ccdnw.musliceQu}
for one transect in $\toyund-\toydir$
parameter space ($\toydir=0.66$, $\toyund$ varying),
and some example values of $\infparam$.

Note that for $\infparam=1$, all nodes
are vulnerable providing they can be reached
along an edge, and the contagion model's behavior exhibits an
inherent symmetry in the network's giant in-component
and out-component, implying that $\Ptrig = \phi_\infty$.
For the infection probability
$\Ptrig$, the initial node must be part of the 
giant in-component which is made up of
type 1, 2, and 4 nodes.  The final infected component
will match the giant out-component which is turn
made up of type 1, 3, and 4 nodes.
The giant strongly connected component is found
in the intersection: type 1 and 4 nodes.

\subsection{Final Size of Infection}
\label{subsec:ccdnw.toymodel-finalsize}

\begin{figure}[tp!]
  \centering
  \includegraphics[width=0.49\textwidth]{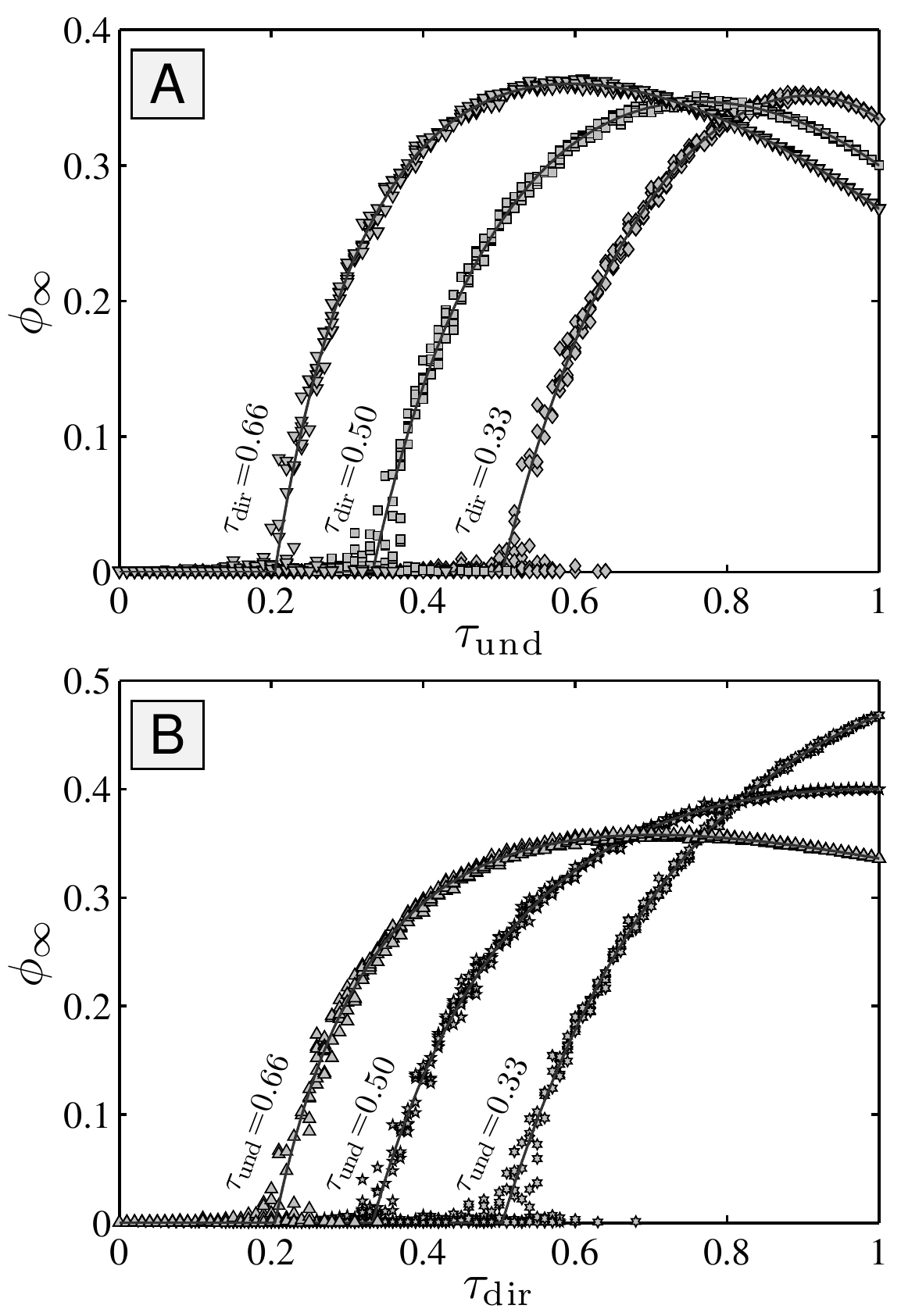}
        \caption{
    Final size curves for example parameter choices
    for the toy model described
    in Sec.~\ref{sec:ccdnw.toymodel}
    with $\infparam=1$.
    Symbols indicate sample output from simulations with $N=10^5$
    and solid curves follow from
    Eqs.\req{eq:ccdnw.phifinalsize},
    \req{eq:ccdnw.y-sol1},
    \req{eq:ccdnw.x},
    \req{eq:ccdnw.u},
    and
    \req{eq:ccdnw.v}.
    For each value of $\toyund$ and $\toydir$ in
    the plots A and B,
    we show
    a maximum of 10 values of $\phi_{\infty}$ randomly chosen from $10^5$
    individual simulations 
    for which the final size exceeds $\phi_{\infty} > 50$ (or 0.05\%).
    The overall fit between theory and simulation is excellent.
    See the captions of Fig.~\ref{fig:ccdnw.schematic} 
    and Fig.~\ref{fig:ccdnw.musliceQu} for
    further simulation details.
  }
  \label{fig:ccdnw.muslice}
\end{figure}

We use the evolution equations
given in Sec.~\ref{sec:ccdnw.finalsize}
to describe the growth of the spreading process
on our example networks, with the main
goal of determining the final size.
We start with the equations for edge infection
probabilities,
$\theta_{\veck,t+1}^{(\bidmark)}$
and
$\theta_{\veck,t+1}^{(\outmark)}$,
Eqs.~\req{eq:ccdnw.edgeEvU} and~\req{eq:ccdnw.edgeEvI},
and we present their full model-specific forms
in the Appendix.
We let $t \to \infty$ in these equations
and solve for fixed points,
$\theta_{\veck,\infty}^{(\bidmark)}$
and
$\theta_{\veck,\infty}^{(\outmark)}$.
Only two of the eight possible equations 
are coupled, those for
$\theta_{1,\infty}^{(\inmark)}$ and
$\theta_{1,\infty}^{(\bidmark)}$ 
(Eqs.~\ref{eq:ccdnw.theta1u} and~\ref{eq:ccdnw.theta1i}),
and thus they alone determine the final probabilities.
As per our previous calculations, 
this collapse in equation number
is because $\veck_1$ nodes are the only type 
capable of receiving and transmitting an infection.
With these observations, and in setting $\infparam=1$
for simplicity, we obtain the coupled equations
\begin{equation}
  \label{eq:ccdnw.y1}
  \tinfbid = \toyund 
  \left[ 
    \tinfbid 
    \left(
      1-\tinfin
    \right) 
    + \tinfin 
  \right],
\end{equation}
and
\begin{equation}
  \label{eq:ccdnw.x1}
  \tinfin = \toydir 
  \left( 
    \left[
      2\tinfbid - 
      \left(
        \tinfbid
      \right)^2
    \right]
    \left[
      1-\tinfin
    \right]
    + \tinfin 
  \right). 
\end{equation}
Solving both equations for $\tinfin$ and equating the results leads to a
cubic polynomial in $\tinfbid$. One root is $\tinfbid=0$ and the others
are solutions to
\begin{gather}
  \label{eq:ccdnw.exquad}
  0 = 
  \toydir 
  \left(
    \tinfbid
  \right)^2 - 
  \toydir 
  \left(
    \toyund+2
  \right) 
  \tinfbid 
  \nonumber \\
  + 
  \left(\toydir+1\right)
  \left(\toyund+1\right) 
  - 2.
\end{gather}
We look for solutions 
for which $0 \leq \tinfbid \leq 1$.
If 
$
\left(\toydir+1\right)\left(\toyund+1\right) \leq 2
$,
we find the only feasible solution is
$
\tinfbid = 
0$.
When $(\toydir+1)(\toyund+1) > 2$ 
we find $\tinfbid = 0$ again but
now also a non-trivial solution:
\begin{equation}
  \tinfbid = 
  \frac{1}{2} (\toyund+2)
  \left[
    1 - 
    \sqrt{
      1 -
      4\frac{
        (\toydir + 1)(\toyund + 1) - 2
      }
      {
        \toydir^2 \left( \toyund + 2 \right)^2 
      }
    }
    \;
  \right].
  \label{eq:ccdnw.y-sol1}
\end{equation}
As expected, the probability of an infected edge $\tinfbid$ becomes nonzero
as we move away from the phase transition
curve in $(\toydir, \toyund)$-space, 
in agreement with our global spreading condition
analysis, \Req{eq:ccdnw.mulambdacurve}.
Using our expression for $\tinfbid$, we obtain 
expressions for the other non-zero edge infected
probabilities:
\begin{equation}
  \tinfin 
  = 
  \frac{
    (1-\toyund) \tinfbid
  }
  {\toyund 
    \left(
      1-\tinfbid
    \right)
  },
  \label{eq:ccdnw.x}
\end{equation}
\begin{eqnarray}
  \label{eq:ccdnw.u}
  \lefteqn{\theta_{\veck_3,\infty}^{(\inmark)} = (1-\toydir)
    \left[ 
      \tinfin + 
      2\tinfbid 
    \right.
    }
    \\
    &
    \left.
      - 2\tinfin\tinfbid -
    \left(
      \tinfbid
    \right)^2 +
    \left(
      \tinfin\tinfbid
    \right)^2
  \right], \nonumber 
\end{eqnarray}
and
\begin{equation}
  \theta_{\veck_4,\infty}^{(\bidmark)} 
  = 
  \left(
    1-\toyund
  \right)
  \left(
    \tinfbid + \tinfin - \tinfin\tinfbid
  \right). 
  \label{eq:ccdnw.v}
\end{equation}
Our last step is to use the above
edge infection probabilities to compute
the eventual fractional extent of a
global spreading event $\phi_{\infty}$
using~\Req{eq:ccdnw.phifinalsize}
(with $\phi_0 \rightarrow 0$).
In Fig.~\ref{fig:ccdnw.muslice},
we compare output of our simulations
with the model's version of~\Req{eq:ccdnw.phifinalsize},
once again showing excellent agreement.

\section{Concluding remarks}
\label{sec:ccdnw.conclusion}

We have provided an extensive
treatment of spreading on generalized random networks,
accommodating a wide range of contagion processes
from biological to social in nature.
Our analysis is straightforward in that physical intuition
is always at hand, and in no place have we resorted to 
more mathematical, less transparent approaches,
such as those employing generating functions.

In closing, we note that if nodes are capable of
recovery and reinfection, general calculations become
considerably more difficult, particularly regarding
the final extent of a spreading event, 
and this remains an open area of investigation.

\acknowledgments
JLP was supported by NIH grant \# K25-CA134286;
KDH was supported by VT-NASA EPSCoR;
PSD was supported by NSF CAREER Award \# 0846668.
The authors are grateful for the computational 
resources provided by the Vermont Advanced Computing 
Center which is supported by NASA (NNX 08A096G).

\smallskip

\appendix

\section{Final size calculations}

For the model described in Sec.~\ref{sec:ccdnw.toymodel},
we provide the specific forms below for 
Eqs.~\req{eq:ccdnw.edgeEvU} and~\req{eq:ccdnw.edgeEvI} 
as used for the calculations in
Sec.~\ref{subsec:ccdnw.toymodel-finalsize}
regarding the final size of spreading events.

\begin{widetext}
  \begin{multline}
    \theta_{\veck_1,t+1}^{\left(\bidmark\right)} = 
    \phi_0 + \left(1-\phi_0\right)
    \left\{
      \toyund 
      \left[
        \left(1-\theta_{\veck_1,t}^{\left(\inmark\right)}\right)^2 
        \rspfn_{0, \veck_1} +
        \left(1-\theta_{\veck_1,t}^{\left(\inmark\right)}\right) 
        \theta_{\veck_1,t}^{\left(\bidmark\right)} \rspfn_{1,\veck_1} +
      \right.
    \right.
    \\
    \left.
      \left.
        \theta_{\veck_1,t}^{\left(\inmark\right)} \left(1-\theta_{\veck_1,t}^{\left(\bidmark\right)}\right) \rspfn_{1,\veck_1} +
        \theta_{\veck_1,t}^{\left(\inmark\right)} \theta_{\veck_1,t}^{\left(\bidmark\right)} \rspfn_{2, \veck_1}
      \right] 
      + \left(1-\toyund\right) \rspfn_{0, \veck_4}
    \right\}
    \label{eq:ccdnw.theta1u}
  \end{multline}
  \begin{multline}
    \theta_{\veck_1,t+1}^{\left(\inmark\right)} = 
    \phi_0 + \left(1-\phi_0\right)
    \left\{
      \toydir
      \left[
        \left(1-\theta_{\veck_1,t}^{\left(\inmark\right)}\right)
        \left(1-\theta_{\veck_1,t}^{\left(\bidmark\right)}\right)^2 
        \rspfn_{0, \veck_1} +
    \left(1-\theta_{\veck_1,t}^{\left(\inmark\right)}\right) 
    \binom{2}{1} \theta_{\veck_1,t}^{\left(\bidmark\right)} 
    \left(1-\theta_{\veck_1,t}^{\left(\bidmark\right)}\right)
    \rspfn_{1,\veck_1} + 
      \right.
    \right.
    \\
    \left(1-\theta_{\veck_1,t}^{\left(\inmark\right)}\right)
    \left(\theta_{\veck_1,t}^{\left(\bidmark\right)}\right)^2 
    \rspfn_{2, \veck_1} +
    \theta_{\veck_1,t}^{\left(\inmark\right)}
    \binom{2}{1} \theta_{\veck_1,t}^{\left(\bidmark\right)}
    \left(1-\theta_{\veck_1,t}^{\left(\bidmark\right)}\right) 
    \rspfn_{2,\veck_1} +
    \\
    \left.
      \left.
        \theta_{\veck_1,t}^{\left(\inmark\right)}
        \left(\theta_{\veck_1,t}^{\left(\bidmark\right)}\right)^2
        \rspfn_{3,\veck_1}
      \right] 
      + \left(1-\toydir\right) \rspfn_{0, \veck_2}
    \right\}
    \label{eq:ccdnw.theta1i}
  \end{multline}
  \begin{multline}
    \theta_{\veck_3,t+1}^{\left(\inmark\right)} = 
    \phi_0 + \left(1-\phi_0\right)
    \left\{
      \left(1-\toydir\right)
      \left[
        \left(1-\theta_{\veck_1,t}^{\left(\inmark\right)}\right)
        \left(1-\theta_{\veck_1,t}^{\left(\bidmark\right)}\right)^2 
        \rspfn_{0, \veck_1} +
      \right.
    \right.
    \\
    \left(1-\theta_{\veck_1,t}^{\left(\inmark\right)}\right) 
    \binom{2}{1} \theta_{\veck_1,t}^{\left(\bidmark\right)} 
    \left(1-\theta_{\veck_1,t}^{\left(\bidmark\right)}\right)
    \rspfn_{1,\veck_1} + 
    \theta_{\veck_1,t}^{\left(\inmark\right)}
    \left(1-\theta_{\veck_1,t}^{\left(\bidmark\right)}\right)^2 
    \rspfn_{1, \veck_1} +
    \\
    \left(1-\theta_{\veck_1,t}^{\left(\inmark\right)}\right)
    \left(\theta_{\veck_1,t}^{\left(\bidmark\right)}\right)^2 
    \rspfn_{2, \veck_1} +
    \theta_{\veck_1,t}^{\left(\inmark\right)}
    \binom{2}{1} \theta_{\veck_1,t}^{\left(\bidmark\right)}
    \left(1-\theta_{\veck_1,t}^{\left(\bidmark\right)}\right) 
    \rspfn_{2,\veck_1} +
    \\
    \left.
      \left.
        \theta_{\veck_1,t}^{\left(\inmark\right)}
        \left(\theta_{\veck_1,t}^{\left(\bidmark\right)}\right)^2
        \rspfn_{3,\veck_1}
      \right] 
      + \toydir \rspfn_{0, \veck_2}
    \right\}
    \label{eq:ccdnw.theta3i}
  \end{multline}
  \begin{multline}
    \theta_{\veck_4,t+1}^{\left(\bidmark\right)} = 
    \phi_0 + \left(1-\phi_0\right)
    \left\{
      \left(1-\toyund\right)
      \left[
        \left(1-\theta_{\veck_1,t}^{\left(\inmark\right)}\right)^2 \rspfn_{0, \veck_1} +
        \left(1-\theta_{\veck_1,t}^{\left(\inmark\right)}\right) \theta_{\veck_1,t}^{\left(\bidmark\right)} \rspfn_{1,\veck_1} + 
      \right. 
    \right. \\
    \left.
      \left.
        \theta_{\veck_1,t}^{\left(\inmark\right)} \left(1-\theta_{\veck_1,t}^{\left(\bidmark\right)}\right) \rspfn_{1,\veck_1} +
        \theta_{\veck_1,t}^{\left(\inmark\right)} \theta_{\veck_1,t}^{\left(\bidmark\right)} \rspfn_{2, \veck_1}
      \right] 
      + \toyund \rspfn_{0, \veck_4}
    \right\}
    \label{eq:ccdnw.theta4u}
  \end{multline}
\end{widetext}
\begin{equation}
  \theta_{\veck_2}^{\left(\inmark\right)} = \phi_0
  \label{eq:ccdnw.theta2i}
\end{equation}
\begin{equation}
  \theta_{\veck_4}^{\left(\inmark\right)} = \phi_0
  \label{eq:ccdnw.theta4i}
\end{equation}

\end{document}